\documentclass[11pt]{article}
\usepackage{amsmath,amssymb,epsfig,cite}
\advance\hoffset by -1.1truecm
\advance\voffset by -1.0truecm
\newcommand{\be}{\begin{equation}}
\newcommand{\ee}{\end{equation}}
\newcommand{\bea}{\begin{eqnarray}}
\newcommand{\eea}{\end{eqnarray}}

\numberwithin{equation}{section}

\newcounter{appendice}

\begin{document}

\title{\begin{flushright}
 \small SU-4252-861\\IISc/CHEP/12/07
\end{flushright}
\vspace{0.25cm} $S$-Matrix on the Moyal Plane: Locality versus Lorentz
Invariance}
\author{A. P. Balachandran$^a$\footnote{bal@phy.syr.edu} ,
A. Pinzul$^b$\footnote{apinzul@fma.if.usp.br} ,
  B. A. Qureshi$^a$\footnote{bqureshi@phy.syr.edu}
  and S. Vaidya$^c$\footnote{vaidya@cts.iisc.ernet.in}\\ \\
$^a$\begin{small}Department of Physics, Syracuse University, Syracuse NY,
13244-1130, USA. \end{small} \\
$^b$\begin{small}Instituto de F\'{i}sica, Universidade de S\~{a}o
  Paulo, C.P. 66318, S\~{a}o Paulo, SP, 05315-970, Brazil. \end{small} \\
$^c$\begin{small}Centre for High Energy Physics, Indian
Institute of Science,
Bangalore, 560012, India.
\end{small}}
\date{\empty}

\maketitle
\begin{abstract}
Twisted quantum field theories on the GM plane are known to be
non-local. Despite this non-locality, it is possible to define a
generalized notion of causality. We show that interacting quantum
field theories that involve only couplings between matter fields, or
between matter fields and minimally coupled $U(1)$ gauge fields are
causal in this sense. On the other hand, interactions between matter
fields and non-abelian gauge fields violate this generalized
causality. We derive the modified Feynman rules emergent from these
features. They imply that interactions of matter with non-abelian gauge
fields are not Lorentz- and $CPT$-invariant.

\end{abstract}

\section{Introduction}

Quantum field theories on the Groenewold-Moyal (GM) plane can be made
Poincar\'e covariant, provided their statistics are twisted along with
the coproduct on the Poincar\'e group \cite{bmpv,replyto}. It is also
possible to write interacting quantum field theories including gauge
theories, and discuss scattering amplitudes. Such models are unitary
as long as the interaction Hamiltonian is hermitian.

However, twisted quantum fields are also non-local \cite{replyto}.
Naively, this might suggest that the scattering matrix for these
theories cannot be Lorentz-invariant. In this article, we will show
that for a large class of noncommutative field theories, the
$S$-matrix is indeed Lorentz-invariant because of the presence of a
weakened form of locality. (The connection between locality and
Lorentz-invariance of the $S$-matrix for noncommutative theories has
also been noticed by \cite{ccgm}) We will also show that
noncommutative non-abelian gauge theories with matter field
interactions violate even this weakened notion of locality, as a
result of which the $S$-matrix in these theories is not Lorentz
invariant (They also violate $CPT$ \cite{abjj}).

It is not difficult to understand the origin of such
non-invariance. The density $H_I$ of the interaction Hamiltonian is
not a local field when $\theta^{\mu\nu}\neq 0$ in the sense that
\begin{equation}
[H_I (x), H_I (y)] \neq 0, \quad x \sim y \label{Bcausal}
\end{equation}
where $x \sim y$ means that $x$ and $y$ are space-like separated.  But
$S$ involves time-ordered products of $H_I$ and the equality sign in
(\ref{Bcausal}) is used to prove its Lorentz invariance already when
$\theta^{\mu\nu}=0$. This condition on $H_I$, known as Bogoliubov
causality \cite{bog}, has been reviewed and refined by Weinberg
\cite{weinberg1,weinbergbook1}. For $\theta^{\mu\nu}\neq 0$, a certain
generalization of this condition is sufficient for Lorentz
invariance. It is fulfilled in the absence of non-abelian gauge
fields, but is violated in the presence of the latter if non-singlet
matter fields are also present. The nonperturbative LSZ formalism
\cite{weinbergbook1} also leads to the time-ordered product of
relatively non-local fields and is not compatible with Lorentz
invariance for $\theta^{\mu\nu}\neq 0$ and matter-non-abelian gauge
field interactions. Such a breakdown of Lorentz invariance is very
controlled and may provide unique signals for non-commutative
spacetimes, a point which requires further study.

In Section 2, we show that these noncommutative theories without gauge
interactions obey a weaker form the the condition
(\ref{Bcausal}). Consequently, the $S$-matrix of such theories is
Lorentz-invariant. In Section 3, we remark that this feature is
maintained in the presence of just abelian gauge fields. Next we
discuss noncommutative non-abelian gauge theories with non-singlet
matter fields, and show that we lose even this generalized notion of
locality. As a result, the Lorentz invariance of the $S$-matrix is
lost at the quantum level.

As an application of these ideas, we will derive the Feynman rules for
noncommutative $QCD$ (as a specific example) and identify specific
diagrams that violate Lorentz invariance in Sections 3 and 4. The
Pauli principle is not violated by the $S$-matrix for scattering of
particles of definite momenta, as we also discuss.

The phenomenology of such Lorentz and $CPT$ violations remains to be
studied. 

\section{Locality and Lorentz Invariance}

For the purposes of our discussion, locality (causality) will have the
meaning it takes in standard local quantum field theories. Thus if
$\rho(\xi)$ is an observable local field $\rho$ like the electric
charge density localized at a spacetime point $\xi$, and $x$ and $y$
are spacelike separated points ($x \sim y$), then causality (locality)
states that
\begin{equation}
[\rho(x),\rho(y)]=0.
\label{causality}
\end{equation}
It means that $\rho(x)$ and $\rho(y)$ are simultaneously measurable.

Causal set theory (see for example \cite{causalreview} for a recent
review) uses a sense of causality which differs from
(\ref{causality}). There is also a criticism of the conceptual
foundations of (\ref{causality}) by Sorkin \cite{sorkin}.

Let $H_I$ be the interaction Hamiltonian density in the interaction
representation. The interaction representation $S$-matrix is
\begin{equation}
S = T \exp \left( -i \int d^N x H_I(x) \right)\, . \label{Sformal}
\end{equation}
For commutative spacetimes, Bogoliubov and Shirkov \cite{bog} long
ago deduced from causality and relativistic invariance that $H_I$ is
a local field:
\begin{equation}
[H_I(x), H_I(y)]=0, \quad x \sim y \, .
\label{localHam}
\end{equation}

Later Weinberg \cite{weinberg1,weinbergbook1} discussed the
fundamental significance of (\ref{localHam}) for these spacetimes: if
(\ref{localHam}) fails, then $S$ is not relativistically invariant.

In these previous discussions, where $\theta^{\mu \nu}=0$, $H_I$ and
their products were taken to transform in the standard way under
Lorentz transformations $\Lambda$:
\begin{eqnarray}
U(\Lambda) H_I(x) &=& H_I(\Lambda^{-1}x)U(\Lambda), \label{scalarH}\\
U(\Lambda)H_I(x)H_I(y) &=& H_I(\Lambda^{-1}x) H_I(\Lambda^{-1}y)
U(\Lambda), \quad {\rm etc}.
\end{eqnarray}

For $\theta^{\mu \nu} \neq 0$, the Lorentz transformation condition on
$H_I$ reduces to (\ref{scalarH}) in the first order term of
(\ref{Sformal}), as our previous work shows \cite{replyto}, and as we
explain later in this section.

However, we must use the twisted coproduct to transform tensor
products of $H_I$. For this twisted coproduct as well, causality or
rather a certain simple generalization of it, is essentially adequate
to guarantee the Lorentz invariance of the $S$-matrix. The
generalization allows for causality, but allows also for weaker
possibilities. It is only ``essentially'' adequate: as Weinberg has
shown \cite{weinberg1}, for a Lorentz-invariant $S$-matrix, there are
also conditions on singularities supported at $x=y$ in the product
$H_I(x)H_I(y)$.

Let us show these results.

i) {\it Lorentz Transformation Law for the S-matrix}

The second order term in (\ref{Sformal}) is the leading term
influenced by time-ordering. It is
\begin{eqnarray}
S^{(2)} &=& \frac{(-i)^2}{2!} \int d^Nx d^Ny \,T(H_I(x) H_I(y)), \\
T(H_I(x) H_I(y)) &=& \theta(x_0 - y_0) H_I(x) H_I(y) + (x
\leftrightarrow y) \label{Torder}\, .
\end{eqnarray}

Thus $S^{(2)}$ is the sum of two terms $S_1^{(2)}$ and $S_1^{(2)}$
corresponding to terms in (\ref{Torder}):
\begin{equation}
S^{(2)}=S_1^{(2)}+S_2^{(2)}.
\end{equation}
In terms of the Fourier transforms $\tilde{H}_I$ of $H_I$,
\begin{equation}
\tilde{H}_I(p)= \int\frac{d^Nx}{(2\pi)^N}e^{ip\cdot x} H_I(x),
\end{equation}
$S_1^{(2)}$ has the expression

\begin{equation}
S_1^{(2)} = -\frac{1}{2} \int \frac{d^N x}{(2\pi)^N}\frac{d^N
  y}{(2\pi)^N}\theta(x^0 - y^0) \int d^N k_1 d^N k_2 \tilde{H}_I(k_1)
  \tilde{H}_I(k_2)e_{k_1}(x) e_{k_2}(y) \, .
\end{equation}

Elsewhere \cite{replyto}, we worked out the twisted transformation
of $e_{k_1} \otimes e_{k_2}$ under $U(\Lambda)$:
\begin{eqnarray}
U(\Lambda)e_{k_1} \otimes e_{k_2} &=& e_{\Lambda k_1} \otimes e_{\Lambda
  k_2}e^{\frac{i}{2} k_1 \cdot \delta_\Lambda \theta \cdot k_2}
  U(\Lambda_2) , \\
\Lambda_2 &=& e^{-\frac{1}{2}(\Lambda k_1 + \Lambda k_2)_\mu
  \theta^{\mu \nu} \partial_\nu} \Lambda e^{\frac{1}{2}(k_1 + k_2)_\mu
  \theta^{\mu \nu} \partial_\nu} , \\
\delta_\Lambda \theta &\equiv& \Lambda^{-1}\theta \Lambda -
  \theta, \quad k_1 \cdot \delta_\Lambda \cdot k_2 \equiv k_{1\mu}
  (\delta_\Lambda \theta)^{\mu\nu}k_{2\nu} \, .
\end{eqnarray}
We can hence write
\begin{eqnarray}
&&U(\Lambda) S_1^{(2)} = \nonumber \\ 
&&-\frac{1}{2} \int d^N k_1 d^N k_2
\tilde{H}_I(k_1) \tilde{H}_I(k_2) \int \frac{d^N x}{(2\pi)^N}
\frac{d^N y}{(2\pi)^N} \theta(x^0 - y^0)
e^{-\frac{i}{2}\frac{\partial}{\partial x^\mu}
\left[\Lambda \delta_\Lambda \theta \Lambda^{-1}\right]^{\mu\nu}
\frac{\partial}{\partial y^\nu}}
  \times \nonumber \\
&& \left(e_{\Lambda k_1}(x) e_{\Lambda
k_2}(y)\left[e^{-\frac{i}{2}\left(
  \frac{\overleftarrow{\partial}}{\partial x^\mu} +
  \frac{\overleftarrow{\partial}}{\partial y^\mu}\right)\theta^{\mu
  \nu} \partial_\nu} U(\Lambda) e^{\frac{i}{2}
      (\Lambda^{-1})_\mu^{\phantom{\mu}\sigma} \left(
  \frac{\overleftarrow{\partial}}{\partial x^\sigma} +
  \frac{\overleftarrow{\partial}}{\partial y^\sigma}\right)\theta^{\mu
  \nu} \partial_\nu}\right]\right)
\end{eqnarray}
where the derivatives do not act on $\theta(x^0-y^0)$.

Now we note certain simple, but important facts:

i) Since
\begin{equation}
\theta(x^0-y^0)\left(\frac{\overleftarrow{\partial}}{\partial
x}+\frac{\overleftarrow{\partial}}{\partial y}\right)_{\mu}=0,
\label{thetader}
\end{equation}
we can let $\left(\frac{\partial}{\partial
x}+\frac{\partial}{\partial y}\right)_{\nu,\mu}$ to act on
$\theta(x^0-y^0)$ as well.

ii) The expression $\frac{\partial}{\partial x^\mu}(\Lambda
  \delta_\Lambda \theta \Lambda^{-1})_{\mu\nu}\frac{\partial}{\partial
  y^\nu}$ gives zero when applied to $\theta(x^0 - y^0)$ because of
  the antisymmetry of $(\Lambda \delta_\Lambda \theta \Lambda^{-1})$:
\begin{eqnarray}  
&&\frac{\partial}{\partial x^\mu}(\Lambda
  \delta_\Lambda \theta \Lambda^{-1})_{\mu\nu}\frac{\partial}{\partial
  y^\nu}\theta(x^0 - y^0) = \nonumber \\
&&\left(\frac{\partial}{\partial x^0}(\Lambda
  \delta_\Lambda \theta \Lambda^{-1})_{0 i}\frac{\partial}{\partial
  y^i} + \frac{\partial}{\partial x^i}(\Lambda
  \delta_\Lambda \theta \Lambda^{-1})_{i 0}\frac{\partial}{\partial
  y^0}\right)\theta(x^0-y^0) = 0.
\end{eqnarray} 
Hence it too can be permitted to act on $\theta(x^0-y^0)$.

Each term in the expression
\begin{equation}
\hat{f}(x,y)= \frac{\partial}{\partial x^\mu} \left(\Lambda
\delta_\Lambda \theta\Lambda^{-1}\right)^{\mu\nu}
\frac{\partial}{\partial y^\nu}
\end{equation}
contains at least one spatial derivative. In particular only the
following terms have time derivatives:
\begin{equation*}
\frac{\partial}{\partial x^0} \left(\Lambda \delta_\Lambda \theta
\Lambda^{-1}\right)^{0 i}\frac{\partial}{\partial y^i}\ +\
\frac{\partial}{\partial x^i}
\left(\Lambda\delta_\lambda\theta\Lambda^{-1}\right)^{i
0}\frac{\partial}{\partial y^0}\,.
\end{equation*}
Thus suppose we encounter a term like the following:
\begin{equation}
\chi(x,y)=\theta(x^0-y^0)\
\left[\hat{f}(x,y)\alpha_1(x)\alpha_2(y)\right]
\end{equation}
where $\hat{f}(x,y)$ acts only on $\alpha_i$. Then it  is a total
spatial divergence:
\begin{eqnarray}
\chi(x,y) &=& \frac{\partial}{\partial x^i}
\left( (\Lambda\delta_\lambda\theta\Lambda^{-1})^{i
j}\frac{\partial}{\partial y^j}\left[\theta(x^0-y^0) \alpha_1(x)
  \alpha_2(y)\right]\right) \nonumber \\
&+&\frac{\partial}{\partial y^i}
\left( (\Lambda\delta_\lambda\theta\Lambda^{-1})^{0i} \theta(x^0-y^0)
\frac{\partial \alpha_1(x)}{\partial x^0} \alpha_2(y)\right) \nonumber \\
&+&\frac{\partial}{\partial x^i} \left(\Lambda \delta_\Lambda
\theta \Lambda^{-1}\right)^{i0} \theta(x^0-y^0)\left[\alpha_1(x)
  \frac{\partial\alpha_2(y)}{\partial y^0}\right] .
\end{eqnarray}
Here time derivatives do not act on $\theta(x^0-y^0)$.

It follows that
\begin{equation}
\int d^4xd^4y\ \chi(x,y)=0
\end{equation}
and hence that
\begin{eqnarray}
&&\int d^4x\, d^4y\, \theta(x^0-y^0) \left[ e^{-\frac{i}{2}\hat{f}(x,y)}
\alpha_1(x) \alpha_2(y) \right] \nonumber \\
&=&\int d^4x\, d^4y\, \theta(x^0-y^0)
\alpha_1(x)\alpha_2(y).
\end{eqnarray}

This identity incidentally easily generalizes to the following sort of
identity as well:
\begin{eqnarray}
&&\int \prod_{i=1}^N d^4x_i\,
Te^{-\frac{i}{2}\left\{\hat{f}(x_1,x_2) + \hat{f}(x_2,x_3) + \cdots +
  \hat{f}(x_{n-1},x_n) \right\}}
\alpha_1(x_1)\alpha_2(x_2)\cdots\alpha_N(x_N) \nonumber \\
&=& \int \prod_{i=1}^N
d^4x_i\,T\left[\alpha_1(x_1)\alpha_2(x_2)\cdots\alpha_N(x_N)\right]\
.\label{id1} 
\end{eqnarray}
Here in the left-hand side, the $\hat{f}$'s do not act on the step
functions in time-ordering.

We can hence write
\begin{eqnarray}
U(\Lambda)S_1^{(2)} &=& -\frac{1}{2}\int d^N k_1\,d^N
k_2\,\tilde{H}_I(k_1)\tilde{H}_I(k_2) \int\frac{d^N
  x}{(2\pi)^N}\frac{d^N y}{(2\pi)^N} \nonumber \\
&& \left[\theta(x^0-y^0) e_{\Lambda k_1}
  (x) e_{\Lambda k_2} (y)\right]
e^{-\frac{i}{2}\left(\frac{\overleftarrow{\partial}}{\partial 
x^\mu}+\frac{\overleftarrow{\partial}}{\partial y^\mu}\right)
  \theta^{\mu\nu} \partial_\nu} \times \nonumber \\
&& \times U(\Lambda) e^{\frac{i}{2}
  (\Lambda^{-1})^\sigma_{\phantom{\sigma}\mu}
  \left(\frac{\overleftarrow{\partial}}{\partial x^\sigma} +
  \frac{\overleftarrow{\partial}}{\partial
    y^\sigma}\right)\theta^{\mu\nu}\partial_\nu}\,.
\end{eqnarray}
We now expand the exponentials, integrate term by term and throw
away surface terms. A similar calculation can be done for
$U(\Lambda)S_2^{(2)}$ as well. We thus finally find,
\begin{eqnarray}
U(\Lambda)S^{(2)}U(\Lambda)^{-1} &=& -\frac{1}{2}\int d^N x\,d^N y\,
T \left( H_I(\Lambda^{-1}x)H_I(\Lambda^{-1}y)\right) \nonumber \\
&=& -\frac{1}{2}\int d^Nx\,d^Ny\,\left\{ \theta\left( (\Lambda
x)^0-(\Lambda y)^0\right)H_I(x)H_I(y)+\, x\leftrightarrow y\right\}
\end{eqnarray}
just as for $\theta^{\mu\nu}=0$.

As such a calculation extends to all orders in $H_I$, we have
\begin{equation}
U(\Lambda)SU(\Lambda)^{-1}= T\ exp\left(-i\int
d^Nx\,H_I(\Lambda^{-1}x)\right).
\label{newS}
\end{equation}

If $x$ and $y$ are time-like separated, then time-ordering is
invariant under $\Lambda \in {\cal L}_+^{\uparrow}$ (and parity):
$\theta((\Lambda x)^0 - (\Lambda y)^0) = \theta(x^0 - y^0)$. But
that is not the case if $x$ and $y$ are space-like separated, $x
\sim y$. In a causal theory, the result
\begin{equation}
[H_I(x), H_I(y)]=0, \quad x \sim y
\label{nclocality}
\end{equation}
holds and helps restore Lorentz invariance of $S$ despite time-ordering.
Weinberg \cite{weinberg1,weinbergbook1} can be consulted for a
detailed proof.

The condition (\ref{nclocality}) is only a sufficient condition for
Lorentz invariance, it is not necessary as well. We shall see below
that non-gauge noncommutative theories fulfill a weaker form of
(\ref{nclocality}) and are still Lorentz-invariant.

{\it ii)  Non-Gauge Noncommutative Theories}

The qft's on the GM plane are not local. This is the case even without
gauge fields. Still in the absence of gauge fields, we showed
elsewhere \cite{bpq} that the $S$-operator has no
$\theta$-dependence. Hence it is Lorentz-invariant if its associated
$\theta^{\mu \nu}=0$ theory is.

This result comes about as follows.

Let us consider a spin zero field $\Phi$ for simplicity as in
\cite{bpq}. For $\Phi$, the annihilation operators for momentum $p$
will be denoted by $a_p$. Then using eq. (7.11) of \cite{bpqv1}, we
get $a_p e_p = c_p e_p e^{\frac{1}{2} \overleftarrow{\partial}_\mu
\theta^{\mu\nu} P_\nu}$ (where $c_p$ is the annihilation operator for
$\theta^{\mu \nu}=0$ and $P_\nu$ is the Fock space momentum operator)
so that
\begin{equation}
\Phi(x) = \Phi^{(0)}(x) e^{\frac{1}{2}
  \frac{\overleftarrow{\partial}}{\partial x^\mu} \theta^{\mu\nu} P_\nu} \ .
\label{fieldmap}
\end{equation} 
where $\Phi^{(0)}(x)$ is made of $c_p$'s and $c^\dagger_p$'s.

We must take $\ast$-products of $e_p$'s when evaluating products of
$\Phi$'s at the same point since $e_p \in {\cal A}_\theta({\mathbb
R}^N)$. It becomes the ordinary product when we substitute
(\ref{fieldmap}) as proved in \cite{bpq} and we get for the $\ast$-product
of $n$ $\Phi$'s, 
\begin{equation} 
\Phi(x)* \Phi(x) \cdots * \Phi(x) = (\Phi^{(0)}(x))^n e^{\frac{1}{2}
  \frac{\overleftarrow{\partial}}{\partial x^\mu} \theta^{\mu\nu} P_\nu}
\end{equation} 
($(\Phi^{(0)}(x))^n$ involves only commutative products of functions.)

Thus in the absence of gauge fields,
\begin{equation}
H_I(x) = H_I^{(0)}(x) e^{\frac{1}{2}
  \frac{\overleftarrow{\partial}}{\partial x^\mu} \theta^{\mu\nu}
  P_\nu} \ ,
\label{ncHam}
\end{equation}
$H_I^{(0)}(x)$ being the interaction density for $\theta^{\mu\nu}=0$. 

Notice that
\begin{equation} 
\int d^4x H_I (x) = \int d^4x H^{(0)}_I(x) \ ,
\end{equation} 
because the exponential factor in (\ref{ncHam}) becomes 1 on
integration over $x$. Also, the Lorentz transformation properties of
$H_I$ can be obtained by transforming the operators $H^{(0)}_I$ and
$P_\nu$ in the standard way \cite{replyto,bpq,bpq1}. Hence the Lorentz
transformation property of the left hand side of (\ref{ncHam}) can be
obtained assuming (\ref{scalarH}).

Since
\begin{equation} 
[P_\nu, H_I(y)] = -i \frac{\partial}{\partial y^\nu} H_I (y) \ ,
\end{equation} 
(\ref{ncHam}) gives for example
\begin{equation}
H_I(x)H_I(y) = H_I^{(0)}(x) e^{\frac{1}{2}
  \frac{\overleftarrow{\partial}}{\partial x^\mu} \theta^{\mu\nu}\left(-i
  \frac{\partial}{\partial y_\nu}\right)}H_I^{(0)}(y) e^{\frac{1}{2}\left(
  \frac{\overleftarrow{\partial}}{\partial x^\mu} +
  \frac{\overleftarrow{\partial}}{\partial y^\mu}\right)
  \theta^{\mu\nu} P_\nu} \ . \label{Hprod}
\end{equation}
Hence,
\begin{eqnarray}
T(H_I(x_1)H_I(x_2) \cdots H_I(x_k)) &=& T\left(H_I^{(0)}(x_1) e^{\frac{1}{2}
  \frac{\overleftarrow{\partial}}{\partial x_1^\mu} \theta^{\mu\nu}
  P^{(1)}_\nu} H_I^{(0)}(x_2) e^{\frac{1}{2}
  \frac{\overleftarrow{\partial}}{\partial x_2^\mu} \theta^{\mu\nu}
  P^{(2)}_\nu} \cdots \right. \nonumber \\
&& \cdots \left. H_I^{(0)}(x_k) e^{\frac{1}{2}\left(
  \frac{\overleftarrow{\partial}}{\partial x_1^\mu} + \cdots
  \frac{\overleftarrow{\partial}}{\partial x_k^\mu} \right)
  \theta^{\mu\nu} P_\nu} \right),  \label{torder} \\
P^{(j)}_\mu &=& -i \left( \frac{\partial}{\partial x_{j+1}^\mu} +
  \cdots \frac{\partial}{\partial x_k^\mu} \right), \quad j \leq k-1
\end{eqnarray} 
where the derivatives in (\ref{torder}) do not act on the
step-functions in the definition of the time-ordered product. But we
can let them act on the step functions as well in view of the
discussion from (\ref{thetader}) to (\ref{id1}). [We must adapt it
  only slightly to reach this conclusion.] Then integrating over
$x_i$'s and discarding surface terms as in (\ref{id1}), we find that
$S$ is independent of $\theta^{\mu\nu}$.

This is a fundamental result of \cite{bpq} in proving the absence of
UV-IR mixing in non-gauge noncommutative theories.

In the same way, we can show that $U(\Lambda) S U(\Lambda^{-1})$
given in (\ref{newS}) is independent of $\theta^{\mu\nu}$:
\begin{equation}
U(\Lambda) S U(\Lambda^{-1}) = T\ exp\left(-i\int d^Nx
H_I(\Lambda^{-1}x)\right) := S^{(0)} \ .
\label{Sinv}
\end{equation}
Thus if the $\theta^{\mu\nu}=0$ theory has a causal interaction
Hamiltonian density $H^{(0)}_I$ and the operator product $H^{(0)}_I(x)
H^{(0)}_I(y)$ is not too singular at $x=y$ so that $S^{(0)}$ is
Lorentz invariant, then $S$ is also Lorentz invariant.

{\it iii) Generalized Causality}

We see from (\ref{Sinv}) that the following generalized causality
condition holds in non-gauge theories for any $\theta^{\mu\nu}$: for
some choice of the constant $\lambda$, the operator
\begin{equation}
H_I^{(\lambda)}(x) = H_I(x) e^{-\frac{1+\lambda}{2}
  \frac{\overleftarrow{\partial}}{\partial x^\mu} \theta^{\mu\nu}
  P_\nu}
\label{causalH}
\end{equation}
is local:
\begin{equation}
[H_I^{(\lambda)}(x), H_I^{(\lambda)}(y)] = 0, \quad x \sim y.
\label{localitylambda}
\end{equation}
This is our generalized causality relation. Our arguments show that if
\begin{equation} 
S = T e^{\left(-i\int d^Nx d H_I(x)\right)}
\end{equation} 
and
\begin{equation} 
S^{(\lambda)} = T e^{\left(-i\int d^Nx d H^{(\lambda)}_I(x)\right)},
\end{equation} 
then
\begin{equation}
S = S^{(\lambda)}\ .
\end{equation}
Weinberg's arguments show that $S^{(0)}$ is Lorentz-invariant if
(\ref{nclocality}) holds unless singularities at coincident points
(mentioned before) spoil it. Therefore $S^{(\lambda)}$ will also be Lorentz
-invariant if (\ref{localitylambda}) holds and singularities at coincident
points do not spoil it.

\section{Gauge Theories with Matter Fields}

Suppose we have a charged scalar field $\Phi$,
\begin{equation}
\Phi(x) = \int d\mu(p) (a_p e^{-i p \cdot x} + b^\dagger(p) e^{i p
  \cdot x})
\end{equation}
that obeys twisted statistics. Then $\Phi$ can be written in terms of
the corresponding commutative counterpart $\Phi^{(0)}$ using
(\ref{fieldmap}), where
\begin{equation} 
P_\mu = \int d\mu(q) q_\mu [a^\dagger(q) a(q) + b^\dagger (q) b(q)] =
{\rm the \,\,total \,\, momentum\,\, operator} .
\end{equation} 

As we discussed in Section 7 of \cite{bpqv1}, we require that the
definition of the covariant derivative $D_\mu$ of the field $\Phi$
preserves statistics, transforms covariantly under Poincar\'e
transformations and has the commutator $[D_\mu,D_\nu]$ given by the
curvature $F^c_{\mu\nu}$ of the commutative gauge fields. This
immediately tells us that $D_\mu$ is of the form
\begin{equation}
D_\mu \Phi = (D_\mu^{(0)} \Phi^{(0)})e^{\frac{1}{2}
   \overleftarrow{\partial}_\mu \theta^{\mu\nu} P_\nu}
\end{equation} 
where
\begin{equation}
D_\mu^{(0)} = \partial_\mu + A_\mu^{(0)}
\end{equation}
and $A_\mu^{(0)}$ is the commutative gauge field. This choice satisfies
all our requirements of a covariant derivative. It also obeys gauge
invariance at the quantum level \cite{bpqv1}. Any gauge group can be
treated in this approach, unlike some other approaches. 

Note that since the gauge symmetry generators are the same as those
for $\theta^{\mu\nu}=0$, the $(F^{(0)}_{\mu \nu})^2$ term of the gauge
field ``kinetic energy term'' also transforms correctly.

Similar arguments can be made about the transformation properties
under the Poincar\'{e} group.

The interaction Hamiltonian splits into two parts:
\begin{equation}
H_\theta^{I} = \int d^3 x [{\cal H}_\theta^{MG} + {\cal H}_\theta^G],
  \quad  MG = {\rm matter-gauge}, \quad G = {\rm pure\,\, gauge\,\, field}
\end{equation}
\begin{eqnarray}
{\cal H}_\theta^{MG} &=& {\cal H}_0^{MG} e^{\frac{1}{2}
  \overleftarrow{\partial}_\mu \theta^{\mu\nu} P_\nu}, \\
{\cal H}_\theta^G &=& {\cal H}_0^G \ .
\end{eqnarray}
We include matter-gauge field and pure matter field couplings in
${\cal H}_\theta^{MG}$, while ${\cal H}_\theta^G$ contains only gauge
field terms.

For $QED$, ${\cal H}_\theta^G =0$ and the $S$-operator of the theory is
the same as in the commutative case:
\begin{equation}
S_\theta^{QED} = S_0^{QED} \ .
\label{Sqed}
\end{equation}
[However, in \cite{bpqv0}, we developed another approach to gauge
  theories where (\ref{Sqed}) is not true.)

For the Standard Model (SM), ${\cal H}_\theta^G = {\cal H}_0^G \neq
0$. As this term has no statistics twist,
\begin{equation}
S_\theta^{SM} \neq S_0^{SM}
\end{equation}
because of the cross-terms in the $S$-matrix between ${\cal
  H}_\theta^{MG}$ and ${\cal H}_\theta^G$. In particular, this inequality
  happens in QCD. Processes like $qg \rightarrow qg$ via a gluon
  exchange interaction actually also violate Lorentz invariance, as we
  explain below.

The generalized causality condition (\ref{localitylambda}) is not
fulfilled in non-abelian gauge theories with matter-gauge field
couplings. It is enough to show this in QCD as we now will.

We have, as in (\ref{fieldmap}),
\begin{equation} 
\Psi(x) = \Psi^{(0)}(x) e^{\frac{1}{2}
  \frac{\overleftarrow{\partial}}{\partial x^\mu} \theta^{\mu\nu}
  P_\nu} \ .
\end{equation} 
$P_{\nu}$ is the {\it total} momentum operator of the quark {\it and}
gluon fields as in (\ref{fieldmap}). That is so for the following
reason. Under covariant transport, $\Psi$ and $D_\mu \Psi$ must have
similar braiding properties. In particular since 
\begin{equation} 
\Psi(x) \Psi(y) = e^{-\frac{i}{2} \frac{\partial}{\partial x^\mu}
  \otimes \theta^{\mu\nu} \frac{\partial}{\partial
  y^\nu}}\left(\Psi^{(0)}(x) \Psi^{(0)}(y)\right)
  e^{\frac{1}{2}\overleftarrow{\partial}_\mu \theta^{\mu\nu} P_\nu}\ ,
\end{equation} 
we need 
\begin{equation} 
D_\mu \Psi(x) D_\nu \Psi(y) = e^{-\frac{i}{2} \frac{\partial}{\partial x^\mu}
  \otimes \theta^{\mu\nu} \frac{\partial}{\partial y^\nu}} \left(D_\mu^{(0)}
  \Psi^{(0)}(x) D_\nu^{(0)} \Psi^{(0)}(y)\right)
  e^{\frac{1}{2}\overleftarrow{\partial}_\mu \theta^{\mu\nu} P_\nu}\ .
\end{equation} 
So this requires 
\begin{equation} 
[P_\mu, D_\lambda \Psi] = -i \partial_\mu D_\lambda \Psi \ .
\end{equation} 
As $D_\lambda$ involves the gluon field, $P_\mu$ must contain its
momentum too. It follows that
\begin{equation} 
H_\theta^I = \frac{e}{2} \left(\bar{\Psi}^{(0)} \gamma \cdot A
\Psi^{(0)}\right) e^{\frac{1}{2} \overleftarrow{\partial}_\mu
  \theta^{\mu\nu} P_\nu} + H_\theta^G 
\end{equation}  
where $H_\theta^G = H_0^G$ contains three- and four-gluon terms and
gluon fields are free. 

As
\begin{equation} 
[P_\mu, H_\theta^G] = -i \partial_\mu H_\theta^G\ ,
\end{equation} 
it is clear that 
\begin{equation} 
[H_\theta^I(x), H_\theta^I(y)] \neq 0\quad {\rm if} \quad x \sim y\ ,
\label{nocausal}
\end{equation}
the non-vanishing term coming from 
\begin{equation} 
[H_\theta^{MG}(x), H_\theta^G(y)]+ x\leftrightarrow y \ .
\end{equation} 
Thus $H_\theta^I$ is not local. It does not fulfill our generalized
locality condition as well. Thus in the next subsection, we explicitly
show that diagrams involving $H_\theta^{MG} H_\theta^G$ lead to
violations of Lorentz invariance in scattering. This proves that
$H_\theta^I$ does not fulfill our generalized causality.

\subsection{Feynman Rules and Examples}

Let $\Psi(x)$ be the noncommutative quantum field representing the
quark. Using (\ref{fieldmap}), it can written in terms of the field
$\Psi^{(0)}(x)$ (with the $\theta^{\mu\nu}=0$ creation-annihilation
operators) as
\begin{equation} 
\Psi(x) = \Psi^{(0)}(x) e^{\frac{1}{2}
  \frac{\overleftarrow{\partial}}{\partial x^\mu} \theta^{\mu\nu}
  P_\nu} \ .
\label{quarkfieldmap} 
\end{equation} 
$P_{\nu}$ is the {\it total} momentum operator of the quark {\it and}
gluon fields as emphasized above. 

Diagrams involving $H_\theta^{MG} H_\theta^G$ lead to violations of
Lorentz invariance in scattering, as we will show below.

The discussion generalizes to the $\theta$-deformed standard model
($SM_\theta$) or any such $\theta$-deformed theory.

In the expansion of $S$, terms involving just $H^{MG}_\theta$ or
just $H^G_\theta \equiv H^G_0$ are independent of $\theta$. The
dependence on $\theta$ comes from terms which involve product
$H^{MG}_\theta$ with $H^G_\theta$. The simplest such term is
\begin{equation}
S^{(2)} = \frac{(-i)^2}{2!} \int d^4x_1 d^4x_2
\,T(H^{MG}_\theta(x_1) H^G_0(x_2)) \ . \label{qg}
\end{equation}
It contributes to quark-gluon $(qg)$ scattering at the tree level,
as shown in Fig.1.

\begin{figure}
\centerline{\epsfig{figure=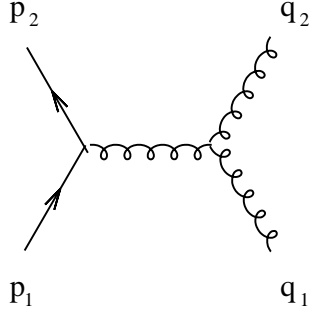,clip=4cm,width=4cm}}
\caption{A Feynman diagram with a non-trivial $\theta$-dependence}
\label{fig:qcd}
\end{figure}

We now simplify $S^{(2)}$. Such simplifications generalize to
arbitrary terms in $S$ as we later indicate.

{\it i) Simplifications for Figure \ref{fig:qcd}}

a) The first simplification comes from integrating over $d^3 x_1$
and throwing away surface terms from spatial derivatives in
$\overleftarrow{\partial}_\mu \theta^{\mu\nu} P_\nu$. This lets us replace
$H^{MG}_\theta(x_1)$ by
\begin{equation}
\widehat{H}^{MG}_\theta(x_1) = H^{MG}_0(x_1)e^{\frac{1}{2}
  \frac{\overleftarrow{\partial}}{\partial x_{10}}
  \theta^{0i}P_i}\ .
\end{equation}

b) We have for $i=1,2,3,$
\begin{equation}
\left[ P_i , \int d^3 x_2  H^G_0(x_2)\right] = 0 \ .
\end{equation}
Hence we can move $P_i$ to the right extreme:
\begin{equation}
S^{(2)} = -\frac{1}{2} \int d^4x_1 d^4x_2 \,T\left(H^{MG}_0(x_1)
H^G_0(x_2)e^{\frac{1}{2}
  \frac{\overleftarrow{\partial}}{\partial x_{10}}
  \theta^{0i}P_i}\right)
\label{s2}
\end{equation}
where $\frac{\overleftarrow{\partial}}{\partial x_{10}}$ does
\textit{not} act the step functions in time defining $T$.

From (\ref{s2}) we see that $P_i$ can be replaced by the total
incident momentum $P_{inc,i}=(p_1 + q_1)_i$ when considering the
process in Figure \ref{fig:qcd}:
\begin{eqnarray}
S^{(2)} &=& -\frac{1}{2} \int d^4x_1 d^4x_2 \,\left\{\theta
(x_{10}-x_{20})\left(H^{MG}_0(x_1)e^{\frac{1}{2}
  \frac{\overleftarrow{\partial}}{\partial x_{10}}
  \vec{\theta}^{0}\cdot\vec{P}_{inc}}  H^G_0(x_2)\right)\right. \nonumber\\
& & + \left. \theta (x_{20}-x_{10})\left(
H^G_0(x_2)H^{MG}_0(x_1)e^{\frac{1}{2}
  \frac{\overleftarrow{\partial}}{\partial x_{10}}
  \vec{\theta}^{0}\cdot\vec{P}_{inc}}\right)\right\}\ ,\label{s21} \\
\vec{\theta}^{0}\cdot\vec{P}_{inc} &=& {\theta}^{0i}{P}_{inc,i}.
\end{eqnarray}

Now
\begin{equation}
H^{MG}_0(x_1)e^{\frac{1}{2}
  \frac{\overleftarrow{\partial}}{\partial x_{10}}
  \vec{\theta}^{0}\cdot\vec{P}_{inc}} = H^{MG}_0(\vec{x}_1, x_{10} +
  \frac{1}{2} \vec{\theta}^{0}\cdot\vec{P}_{inc})\ .
\end{equation}

The $\theta$-deformation thus twists the fields at the $q-q-g$
vertex.

c) By a change of variables, we can shift the deformation to the
$g-g-g$ vertex instead:
\begin{eqnarray}
S^{(2)} &=& -\frac{1}{2} \int d^4x_1 d^4x_2 \,\left\{\theta
(x_{10}-x_{20}-\frac{1}{2}
  \vec{\theta}^{0}\cdot\vec{P}_{inc})\left(H^{MG}_0(x_1)
  H^G_0(x_2)\right)\right. \nonumber\\ 
& & + \left. \theta (x_{20} + \frac{1}{2}
  \vec{\theta}^{0}\cdot\vec{P}_{inc}-x_{10})\left(
H^G_0(x_2)H^{MG}_0(x_1)\right)\right\}=\nonumber\\
& &-\frac{1}{2} \int d^4x_1 d^4x_2 \,T\left(H^{MG}_0(x_1)
H^G_0(\vec{x}_2,x_{20}+\frac{1}{2}
  \vec{\theta}^{0}\cdot\vec{P}_{inc})\right)\ .\label{s22}
\end{eqnarray}
The ability to shift the twist between a quark-quark-gluon and a 3- or
4-gluon vertex connected to it in this manner is often useful. It is
thus sufficient (see also below) to give the twisted gluon
propagator to calculate Feynman diagrams.

{\it ii) The Twisted Gluon Propagator}

The twisted gluon propagator coming from (\ref{s22}) is
\begin{equation}
T\langle A^\alpha_\mu (x_1)A^\beta_\nu
(\vec{x}_2,x_{20}+\frac{1}{2}
  \vec{\theta}^{0}\cdot\vec{P}_{inc})\rangle =
  \delta^{\alpha\beta}\eta_{\mu\nu}D^\theta_F (x_1-x_2)
\end{equation}
where in the Lorentz gauge, $D^\theta_F$ is just the twisted
propagator of a massless scalar field $A$:
\begin{equation}
D^\theta_F (x) = T\langle A(x)A(\vec{0},\frac{1}{2}
  \vec{\theta}^{0}\cdot\vec{P}_{inc})\rangle \ .
\end{equation}
The Fourier expansion of $A$ is
\begin{eqnarray}
A(x) &=& \int \frac{d^3 k}{2k_0}\left( c_k e_k (x) + c^\dagger_k
e_{-k}(x)\right)\ ,\nonumber \\
e_k(x) &=& e^{-ikx}= e^{-i(\vec{k}\cdot\vec{x}-k_0x_0)}\
,\nonumber\\
k_0 &=& |\vec{k}|
\end{eqnarray}
where $c_k , c^\dagger_k$ are the $\theta{\mu\nu}=0$ annihilation
and creation operators. Hence
\begin{eqnarray}
A(\vec{0},\frac{1}{2} \vec{\theta}^{0}\cdot\vec{P}_{inc}) &=& \int
  \frac{d^3 k}{2|\vec{k}|} 
  \left( c_k e^{\frac{i}{2}|\vec{k}|\vec{\theta}^0\cdot\vec{P}_{inc}}
  + c^\dagger_k
  e^{-\frac{i}{2}|\vec{k}|\vec{\theta}^0\cdot\vec{P}_{inc}}\right)\
  \,. 
\end{eqnarray}

Note that we pick up the second term here in the $\theta (x_0)$
term of the $T$-product, and the first term in the $\theta (-x_0)$
term, and these have opposite phases.

Now
\begin{equation}
D^0_F (x) = 2\pi i \int \frac{d^3 k}{2|\vec{k}|}\left(\theta
(x_0)e^{ikx}+\theta (-x_0)e^{-ikx}\right)\ ,
\end{equation}
which comes from
\begin{eqnarray}
D^0_F (x) = \int d^4 k \frac{e^{-ikx}}{k^2 - i\epsilon} &=& -\int
d^3 k e^{-i\vec{k}\cdot\vec{x}}\int \frac{d
k_0}{2|\vec{k}|}e^{ik_0 x_0}\times \nonumber \\
&
&\left(\frac{1}{k_0+|\vec{k}|-i\epsilon} - \frac{1}{k_0-|\vec{k}| +
  i\epsilon}\right)\ .
\end{eqnarray}
Hence
\begin{eqnarray}
D^\theta_F (x) &=& -\int d^3 k e^{-i\vec{k}\cdot\vec{x}}\int
\frac{d k_0}{2|\vec{k}|}e^{ik_0 x_0}\left(
\frac{e^{-\frac{i}{2}|\vec{k}|\vec{\theta}^0\cdot\vec{P}_{inc}}}{k_0 +
  |\vec{k}|-i\epsilon} - \frac{e^{\frac{i}{2}|\vec{k}|\vec{\theta}^0
    \cdot \vec{P}_{inc}}}{k_0-|\vec{k}|+i\epsilon}\right)\nonumber \\
&=& \int d^4 k \frac{e^{-ikx}}{k^2 - i\epsilon}
\left(\cos(|\vec{k}|\vec{\theta}^0\cdot\vec{P}_{inc}) +
i\frac{k_0}{|\vec{k}|}\sin(|\vec{k}|\vec{\theta}^0 \cdot
\vec{P}_{inc})\right)\nonumber\\  
&\equiv& \int d^4 k e^{-i k \cdot x} \widetilde{D}^\theta_F (k)\ .
\end{eqnarray}

{\it iii) General Rules}

In any scattering process, the twist factors
$e^{\frac{1}{2}\overleftarrow{\partial}_\mu \theta^{\mu\nu} P_\nu}$
can all be replaced by
$e^{\frac{1}{2}\overleftarrow{\partial_0}\vec{\theta}^0 \cdot
\vec{P}_{inc}}$ where $\vec{P}_{inc}$ is the incident total momentum
and $\overleftarrow{\partial_0}$ differentiates an appropriate time
argument.

The propagator of a quark or of a gluon connecting two $q-q-g$
vertices is not changed. That is because for example
\begin{eqnarray}
& &\int d^4x_1 d^4x_2 \,\theta
(x_{10}-x_{20})H^{MG}_0(\vec{x}_1,x_{10}+\frac{1}{2}
  \vec{\theta}^{0}\cdot\vec{P}_{inc})
  H^{MG}_0(\vec{x}_2,x_{20} +
  \frac{1}{2}\vec{\theta}^{0}\cdot\vec{P}_{inc})=\nonumber \\
& &\int d^4x_1 d^4x_2 \,\theta (x_{10}-x_{20})H^{MG}_0(x_1)
  H^{MG}_0(x_2)\ .
\end{eqnarray}

In an arbitrary diagram, {\it a priori}, the twisted vertices are the
$q-q-g$ vertices. By a change of variables, we can then shift the
twist to appropriate gluon propagators. In this way, we can tell which
of the gluon propagators in the diagram are twisted.

\section{Lorentz Invariance and Pauli Principle}

{\it i) Violation of Lorentz Invariance}

Consider Fig.1. It carries the propagator
\begin{equation}
\widetilde{D}^\theta_F
(k)\frac{\cos(|\vec{k}|\vec{\theta}^0 \cdot \vec{P}_{inc}) +
  i\frac{k_0}{|\vec{k}|}\sin(|\vec{k}|\vec{\theta}^0 \cdot
  \vec{P}_{inc})}{k^2-i\epsilon}\ .
\end{equation}
The numerator is frame-dependent. It is unity if
\begin{equation}
\vec{\theta}^0\cdot\vec{P}_{inc} = 0\ , \label{com}
\end{equation}
in particular in the center-of mass system. Hence all twist effects
are absent in $S$ in any frame fulfilling (\ref{com}). Otherwise it
depends on $\theta$. Thus as anticipated, the process violates Lorentz
invariance. 

The discussion of $C,P,T$ and $CPT$ can be found in \cite{abjj}.

{\it ii) Pauli Principle Violation}

In \cite{bpqv0}, based on a different treatment of dynamics, we found
Pauli principle violation in processes like electron-electron
scattering. Such violation was present even for cross-sections for
scattering of particles with definite momenta.

In the present approach, there is no such violation in any
scattering cross-section of particles with definite momenta.

But there are expected to be signals of Pauli principle violations
if initial and final particles do not have definite momenta, for
example if they are spatially localized wavepackets. See
for example \cite{bmpv,cghs}.

The proof is very general and very simple too: we just show below that
the initial and final states of definite momenta differ from those for
$\theta_{\mu\nu}=0$ only by a phase, a result well-known. The phase
disappears when we compute cross-sections, that is, in the modulus of
scattering amplitudes. Hence the modulus of scattering amplitudes in
the momentum basis inherits exactly the same symmetry properties from
the states under particle exchange as those for $\theta_{\mu\nu}=0$.
The non-trivial dependence of $S$-matrix on external momenta through
the term $\vec{\theta^0}. \vec{P}_{in}$ does not spoil this argument
because this dependence {\it always} involves the total momentum,
which is, of course, symmetric under permutation of the individual
momenta. We can even replace the actual scattering amplitudes with
ones with the same symmetries under particle exchange as those for
$\theta^{\mu\nu} = 0$ by setting the above-mentioned phase to 1. The
result on Pauli principle follows.

The difference between arbitrary states (such as spatially localized
wave packets) for $\theta_{\mu\nu}=0$ and $\theta_{\mu\nu}\ne 0$ is
not a phase \cite{Balachandran:2007mj}. Hence we cannot readily assert
that the modulus of scattering amplitudes for $\theta_{\mu\nu}=0$ and
$\theta_{\mu\nu}\ne 0$ have the same symmetry under particle exchange
in {\it any} basis.

Now for the demonstration. Consider for example an
\textit{N}-particle state of identical spin-$\frac{1}{2}$
particles. Their creation operators $a_p^{(\lambda)\dagger}$ for
spin basis label $\lambda$ and momentum $p$ are related to those
for $\theta_{\mu\nu}=0$ by
\begin{equation} 
a_p^{(\lambda)\dagger} =
c_p^{(\lambda)\dagger}e^{\frac{i}{2}p\wedge P}, \quad p \wedge P :=
p_\mu \theta^{\mu\nu} P_\nu \ ,
\end{equation} 
where $P$ is the total momentum operator.

For the gauge field, the creation operators $\alpha_q^{(m)\dagger}$
are independent of $\theta_{\mu\nu}$.

Let us first look at a two spin-$\frac{1}{2}$ particle state:
\begin{equation} 
a_{p_1}^{(\lambda_1)\dagger}a_{p_2}^{(\lambda_2)\dagger}|0\rangle =
c_{p_1}^{(\lambda_1)\dagger}c_{p_2}^{(\lambda_2)\dagger}|0\rangle
e^{\frac{i}{2}p_1\wedge p_2}\ .
\end{equation}  
The $\theta$-dependent term on the right side is just a phase. A
similar calculation can be made for any \textit{N} spin-$\frac{1}{2}$
particles and also for any state with bosons, fermions and gauge
particles. Thus in the state
\begin{equation} 
a_{p_1}^{(\lambda_1)\dagger}a_{p_2}^{(\lambda_2)\dagger} \cdots
a_{p_N}^{(\lambda_N)\dagger}\alpha_{q_1}^{(m_1)\dagger}
\alpha_{q_2}^{(m_2)\dagger} \cdots
\alpha_{q_M}^{(m_M)\dagger}|0\rangle \ ,
\end{equation} 
we can move all $P_{\mu}$'s to the right extreme, where they
contribute only a phase. For example, for $N=2$ and $M=1$, the
above expression is
\begin{equation} 
c_{p_1}^{(\lambda_1)\dagger} c_{p_2}^{(\lambda_2)\dagger}
\alpha_{q_1}^{(m_1)\dagger}|0\rangle 
e^{\frac{i}{2}p_1\wedge (p_2+ q_1)}e^{\frac{i}{2}p_2\wedge q_1}\ .
\end{equation} 
In this way, we arrive at our conclusion about Pauli principle.

{\bf Acknowledgments:} It is a pleasure to thank Earnest Akofor, Sang
Jo and Anosh Joseph for discussions. Some of the results of this
papers overlap with some of those of \cite{abjj}. The work of APB and
BQ is supported in part by DOE under grant number DE-FG02-85ER40231. 
The work of AP is supported by FAPESP grant number 06/56056-0.

\end{document}